\documentclass[reqno,11pt]{amsart}

\usepackage{amsmath}
\usepackage{amsfonts}
\usepackage{amssymb}
\usepackage{amsthm}

\usepackage{physics}
\usepackage{tensor}
\usepackage{slashed}

\usepackage{bbold}
\usepackage{mathtools}

\usepackage[nolist]{acronym}
\usepackage{hyperref}
\usepackage{changes}
\usepackage{biblatex}

\usepackage{pgfplots}

\newcommand{\opnorm}[1]{
        {\left| \kern-0.25ex \left| \kern-0.25ex \left| #1 \right| \kern-0.25ex \right| \kern-0.25ex \right|}
}

\newcommand{\R}{\mathbb{R}}
\newcommand{\C}{\mathbb{C}}
\renewcommand{\H}{\mathcal{H}}
\newcommand{\F}{{\mathcal{F}}}

\newtheorem{definition}{Definition}[section]

\newtheorem{example}[definition]{Example}
\newtheorem{conjecture}[definition]{Conjecture}
\newtheorem{remark}[definition]{Remark}
\newtheorem{idea}{Idea}

\numberwithin{equation}{section}

\begin{acronym}
    \acro{cfs}[CFS]{causal fermion system}
    \acro{el}[EL]{Euler-Lagrange}
    \acro{qt}[QT]{quantum theory}
    \acro{qrf}[QRF]{quantum reference frames}
    \acro{gr}[GR]{general relativity}
    \acro{fcdm}[FCDM]{fermionic condensate dark matter}
\end{acronym}

\newcounter{mnotecount}[subsection]

\begin{document}
    \title[QRF's and CG]{Quantum Reference Frames and Correlation Geometry}

    \author[C. F. Paganini]{Claudio F. Paganini$^{*}$}
    \email{claudio.paganini@mathematik.uni-regensburg.de}

    \address{$^*$ Fakultät für Mathematik \\ Universität Regensburg \\ D-93040 Regensburg \\ Germany}

    \begin{abstract}
     The aim of this paper is to provide a largely self-contained, compact and comprehensible introduction to the basic ideas behind correlation geometry, which underlies the theory of \ac{cfs}. A key focus here is on the manner in which the framework deals with gauge transformations, including diffeomorphisms via the principle of unitary equivalence. We will argue that, conceptually, the fundamental description of a physical system in terms of its correlation geometry is much closer to thermodynamics than quantum theory. 
    \end{abstract}

    \maketitle
    \tableofcontents

    \section{Introduction}\label{sec:introduction}
This paper was inspired by a recent paper by Kabel et al.~\cite{kabel2025quantum} as an opportunity to clarify and explain the conceptual foundations of the theory of \ac{cfs}. In particular, we will reiterate the claim in \cite{fischer2025causal} that a superposition of spacetimes widely discussed in the literature, e.g., in the aforementioned  \cite{kabel2025quantum} but also \cite{belenchia2018quantum,zych2019bell,christodoulou2019possibility,anastopoulos2020quantum,giacomini2022quantum,foo2023quantum} and references therein, does not make sense in our setting beyond the weak-field/linearized regime potentially accessible in table top experiments, as discussed, e.g., in~\cite{bose2017spin,marletto2017gravitationally,bose2025massive,carney2019tabletop,huggett2023quantum,martin2023gravity} and references therein.
    
As the arguments in \cite{kabel2025quantum} Kabel et al. serve as an essential backdrop against which we will develop the conceptual clarification for \ac{cfs}, we will start the introduction by recalling the questions they set out to answer in their paper and how they build their argument. A key idea in their paper is the introduction of a notion of \emph{quantum coordinates}. In the introduction to their paper they state that ``a priori there does not exist any preferred way of comparing or identifying the locations of objects across different spacetimes in superposition.'' They are then looking for a way to ``thread'' points across different manifolds. To achieve this goal they then make use of four scalar fields, ``using
their coincidences to construct a comparison map between all spacetime
points in the superposition.'' They refer to these fields as quantum reference frames. They then also discuss gauge fixing procedures that allow for a ``Counter map'' that allows to compare two different physical models. They then proceed to say that ``Thus in general relativity, one needs to ask: What do we mean by
`the same' or a `different' point in spacetime when considering different
solutions to Einstein’s equations? Similarly in electromagnetism: what is
`the same' or a `different' value of the gauge potential across various possible configurations of the electromagnetic field? Usually, we answer
these questions by fixing a coordinate system or gauge across all the different
configurations. More generally, one can do so by fixing a representational
convention.'' 

In the framework of Kabel et al.~\cite{kabel2025quantum} the (quantum) reference frame therefore mainly serves the purpose, to enable you to compare the descriptors of a physical model (i.e. the metric the potentials and so on) across different physical models to then be able to find a meaningful manner to put them into ``superposition''. 

In the following, we will explain how these concepts and ideas are realized in the setting of correlation geometry foundational to the theory of \ac{cfs} and how correlation geometry is essentially the idea of a  \ac{qrf}  taken to its logical conclusion. It is important to note here that the field of \ac{qrf} is not a monolithic theory but a heterogeneous set of ideas; see, e.g.,~\cite{lake2023quantum,goeller2022frames, quantumFrameRelativity, bartlett2007reference, fewster2025quantum,Apadula2024quantumreference, ali2024quantum, Krumm2021quantumreference,Yang2020switchingquantum,carette2025operational,giacomini2019quantum,hardy2020implementation,hardy2018construction,Miyadera_2016,Loveridge_2017,loveridge2018symmetry}. As the present paper was inspired by~\cite{kabel2025quantum}, we will relate the concepts in correlation geometry primarily to their setting. 
The key idea is that instead of using the reference fields (not necessarily scalar) just as labels for the descriptors of the effective physical model, we encode all aspects of the effective physical model directly in the reference fields and their (cor)relations.

The encoding is what we consider to be the fundamental physical model. A key question addressed in \cite{kabel2025quantum} is how to compare different physical models and how to deal with the obstructions that gauge fields pose to this task. Correlation geometry solves both of these problems in an elegant way. As we will demonstrate, there is an unambiguous way to compare different physical models at this fundamental level. Namely, in this paper we will argue that if the correlation geometry of two effective physical models is unitarily equivalent, then the fundamental physical models are the same. This notion in turn provides us with the necessary tools to distinguish which descriptors of the effective physical model are ``physical'' and which are gauge. 

A key conceptual insight from CFS is that we can map an effective physical model expressed in terms of conventional descriptors  (manifold, metric, and matter fields)  to a fundamental physical model expressed in terms of the correlation geometry encoded in (quantum) reference frame fields. The usual descriptors of the effective physical models are then just convenient labels to identify certain special correlation geometries, quite similar to how temperature and pressure in thermodynamics identify a special statistical distribution among all possible statistical distributions. As in thermodynamics, we will argue that \ac{qrf} is more that just a convenient way to describe a physical system, in the sense that it represents the ``true'' fundamental degrees of freedom, analogous to the many-particle picture of a gas, which is indeed a more accurate description of nature.

In the case of thermodynamics it turns out that nature seems to prefer these special distributions, which is the essence of why the macroscopic description of large ensembles of particles in terms of a small number of labels for specific statistical distributions is meaningful at all. In \cite{fischer2025causal} we argued heuristically that the causal action principle exhibits a similar preference for geometries that can be well approximated by a continuum limit, i.e., an effective physical model in terms of conventional descriptors. 

To end the introduction, a quick remark on the language used throughout this paper. 
\begin{itemize}
    \item By a \emph{physical system} , we refer to a real physical object.
    \item By an \emph{effective physical model} , we refer to a description of a physical object valid at some emergent scale. 
    \item By \emph{descriptors} of the effective physical model, we refer to a small number of mathematical objects necessary to characterize the physical system in the effective model. 
    \item By a \emph{fundamental physical model}, we refer to a more elementary description of the physical system that gives rise to the effective physical model in some approximation.
\end{itemize}
As we will frequently use analogies with thermodynamics, we will use thermodynamics to illustrate the use of terminology: The \emph{physical system} in question is a gas in a piston. The \emph{effective physical model} is the description of the gas in thermodynamic language in terms of temperature $T$, volume $V$, and pressure $p$ that together constitute the \emph{descriptors} of this effective physical model. The \emph{fundamental physical model} finally is the description of the gas as a statistical ensemble of particles, mathematically modeled by a density function on phase space. 

% \subsection*{Organization}

\section{The Principles of Correlation Geometry}
In the following, we will sketch the principles of correlation geometry with limited mathematical rigor, since the goal for this paper is not as much to establish a rigorous mathematical framework (this has been done already in~\cite{topology} and will be continued elsewhere) but to communicate the general ideas and to make transparent to what extent this incorporates the conceptual construction at the heart of the arguments in \cite{kabel2025quantum}.  

\begin{idea}[The Classical Starting Point]
    The idea of correlation geometry is to pick a, potentially infinite, set of suitable fields $\{\psi_i\}_{i\in \{1\dots\}}$ defined as sections of the same fiber bundle over the background of the effective physical model to then encode the fundemantal physical model in the correlations between these fields.
\end{idea}

\subsection{The Abstract Definition of the Correlation Geometry}
In this section, we give the abstract definition of the space of all possible correlations  and we will then show how a suitable choice of reference systems allows us to encode effective physical models formulated in terms of conventional descriptors in this one single structure. Having unified the language of an array of effective physical models, it is then straightforward to compare them. Note that the following general definition was given in~\cite{topology} under the name \emph{topological fermion system}\footnote{The motivation to choose a different name here is twofold: First, we believe that it better captures the physics underlying the construction and alignes more closely with the terminology and ideas formulated in the physicy literature, e.g., in~\cite{kempf2021correlational, reitz2023model}. Second, we want to distinguish the mathematical structure (correlation geometry) from the physical theory (causal fermion systems). Hence, one can appreciate the power of the mathematical structure to encode physical information in a novel manner in its own right, without necessarily buying into one specific theory that can be build on it.}.

\begin{definition}[Correlation Geometry]
    A correlation geometry consists of the triplet $(\mathcal{H}, \mathcal{F}^{p,q}, \rho)$ defined as follows:
    \begin{enumerate}
        \item A separable complex Hilbert space $\mathcal{H}$ with a scalar product $(\cdot|\cdot)$.
        \item The subset of the bounded linear operators
        \begin{equation}
            \mathcal{F}^{p,q}:=\{x \in \mathcal{BL}(\mathcal{H})|\text{s.-a. with } x \text{ has max } p \text{ pos. and } q \text{ neg. eigenvalues}\}.
        \end{equation}
        \item A measure $\rho$ defined on the Borel $\sigma$-algebra on $ \mathcal{F}^{p,q}$.
    \end{enumerate}
\end{definition}
The triplet $(\mathcal{H}, \mathcal{F}^{p,q}, \rho)$ will be considered as the fundamental physical model throughout the remainder of this paper. An important fact to point out here is that in \cite[Prop 3.1.3]{Finster2024} it was shown that $\mathcal{F}^{p,q}_{\text{reg}}$ is a smooth manifold of dimension 
\begin{equation}
    \dim (\mathcal{F}^{p,q}_{\text{reg}}) = 2f(p+q)-(p+q)^2
\end{equation}
where $\mathcal{F}^{p,q}_{\text{reg}}$ is the subset of regular elements in $\mathcal{F}^{p,q}$ with exactly $p$ positive and $q$ negative eigenvalues and $f=\dim(\mathcal{H})$ is the dimension of the Hilbert space. This is the underlying reason why we refer to this structure as correlation geometry, with the ``correlation'' part becoming more transparent in the subsequent sections. The set $\mathcal{F}^{p,q}$ can be thought of as the space of all possible correlations of type $p,q$ in a given Hilbert space. 

\begin{remark}
    For $ p=q=n$ we refer to the correlation geometry $(\mathcal{H}, \F^{n,n}, \rho)$ as a causal fermion system of spin dimension $n$. The theory of \ac{cfs} then comes with many additional structures, the most important of which is the causal action principle formulated on $\F^{n,n}$ in terms of the measure $\rho$.
\end{remark}

\subsection{Encoding Manifolds - The Local Correlation Map}\label{sec:encoding}
In the following, we will assume that every (semi) classical physical model $\Phi$ is equipped with a manifold $\mathcal{M}$ as part of its descriptors $\Phi=\Phi(\mathcal{M})$\footnote{The manifold can also be a discrete.}. In fact, to build the correlation geometry, we will assume that the manifold $\mathcal{M}$ is such that it allows for suitable additional structures (e.g. it is differentiable, or it admits a certain fiber bundle).

The goal is now to define a map $F:\Phi\rightarrow(\mathcal{H}, \mathcal{F}^{p,q}, \rho)$ from the effective physical model to its correlation geometry, i.e., its fundamental description. For this, we need the following structures:

\begin{enumerate}
    \item Let $EM$ be a fiber bundle and $\Gamma(EM)$ be the space of smooth sections on the fiber bundle.
    \item  Let $(\cdot |\cdot)_{\tilde{\H}}: S \times S \rightarrow \C (\text{ or }\R)$ be a scalar product on $S \subset\Gamma(EM)$ with the Hilbert space $\tilde{\H}= \overline{S}$ being the closure of $S$ under the scalar product $(\cdot |\cdot)_{\tilde{\H}}$.
    \item  Let $b_x: E_x \times E_x \rightarrow \C$ (or $\R$) be a section of hermitian forms (i.e. $b_x$ evaluated at any point $x\in M$ is a hermitian form on the fiber space). E.g., the inner product $b_x:= \langle\cdot|\cdot \rangle_{E_x}$ could play that role. 
\end{enumerate}

 Then we can identify the local hermitian form with a finite rank operator on the global Hilbert space via the local correlation map
\begin{equation}
    (\psi|\mathbb{F}(x)\psi):= b_x(\psi(x),\phi(x)) \qquad \forall \psi, \phi \in S
\end{equation}
with $\overline{\mathbb{F}(x)}$ being the unique bounded linear extension of $\mathbb{F}(x)$ from $S$ to $\tilde{\H}$. The fact that the local form is hermitian implies that the operator $\overline{\mathbb{F}(x)}$ is self-adjoint. Furthermore, it is clear that since $b_x$ maps from an $\dim(E_x)\times \dim(E_x) $ dimensional space to the real numbers, that the rank of $\overline{\mathbb{F}(x)}$ agrees with the dimension of the fiber space $\text{rank}\left(\overline{\mathbb{F}(x)}\right)= \dim(E_x)$ (if the hermitian form is non-degenerate).

The number of positive and negative eigenvalues of the operator $\overline{\mathbb{F}(x)}$ agrees with the signature of the hermitian form. So, if the hermitian form $b_x$ has signature $(p,q)$, then, upon isometric embedding of $\tilde{\H}\hookrightarrow \H$, the local correlation map 
\begin{align}
   \hphantom{xxxxxxxx} F_S: \qquad &\Phi(M) &\longrightarrow &&\mathcal{F}^{(p,q)} \hphantom{xxxxxxxx}  \\
    & x & \longrightarrow && \mathbb{F}(x)\hphantom{xxxxxxxx} 
\end{align}
maps into the subspace of the bounded linear operators on $\H$. The index $S$ here is a reminder that the local correlation map depends on the subset of smooth sections from which we built our Hilbert space $\tilde{\mathcal{H}}$. Note that in the following, we will always assume an isometric embedding $\tilde{\H}\hookrightarrow \H$ implicitly. 

If the manifold $\mathcal{M}$ of our effective physical model $\Phi$ is equipped with a volume form $\mu$, therefore $\Phi(\mathcal{M},\mu)$, then, given the local correlation map $F_S(\Phi)$, we can define the measure $\rho_S^\Phi$ by the push-forward measure
\begin{equation}
    \rho^\Phi_S(\Omega)=\int_{F_S^{-1}(\Omega)}d\mu \qquad \text{ for all } \qquad \Omega \subset \mathcal{F}^{(p,q)}
\end{equation} 
This completes the description of our physical model $\Phi(\mathcal{M},\mu)$ in terms of its correlation geometry $(\tilde{\mathcal{H}}_S,\mathcal{F}^{(p,q)},  \rho^\Phi_S) $, relative to the subset of smooth sections $S$. In this setting, one should therefore think of $S$, or its completion $\tilde{\H}$ as the (quantum) reference frame for the physical model $\Phi$.

\begin{remark}
    A key difference from the construction of Kabel et al. \cite{kabel2025quantum} is the fact that, in general, one wants to choose $\dim{\H}=N$ as large as possible rather than just consisting of four reference fields. Consider, e.g., the different vectors in $\H$ as plane wave solutions with different wave length. Then the more waves you take into account, the finer the resolution of the local correlation map. If we choose $\dim{\H}=N$ the local correlation map is in a sense ``bandwidth limited'', this notion was discussed in some detail in \cite{fischer2025causal}. The complete, small scale, information of the physical model $\Phi$ can in general only be recovered in the continuum limit where $N\rightarrow \infty$.    
\end{remark}
We demonstrate this problem with the following simple example. For a similar example studied from the perspective of non-smooth geometry and topology, see~\cite{topology}.
\begin{example}\label{ex:one}
    Let our physical model $\Phi(M,g,\mu_g)$ be a compact $n$-dim Riemannian manifold with $\mu_g=\sqrt{|g(x)|}d^n x$. Let $EM=TM$ be the tangent bundle and $S=\{V_1, \dots V_m\} \subset\Gamma(EM)$.  
  We define the Hilbert space as the closure of $S$ via the standard scalar product 
\begin{equation}
    (V_1,V_2)_\H:= \int_M g(x)_{\mu\nu}V_1(x)^\mu V_2(x)^\nu \sqrt{|g(x)|}d^n x \, \qquad \text{ for } V_1, V_2 \in S.
\end{equation} 
and for the local bilinear form we choose
\begin{equation}
    b_x(V_i,V_j)= g(x)_{\mu\nu} V_1(x)^\mu V_2(x)^\nu \, .
\end{equation}
Then the local correlation map 
\begin{equation}
  (V_i,\mathbb{G}(x)V_j)_\H := g(x)_{\mu\nu} V_i(x)^\mu V_j(x)^\nu \qquad \forall V_i,V_j \in S
\end{equation}
 identifies the metric at a point $x$ with a self adjoint rank $n$  operator on the Hilbert space choose
\begin{equation}
  F_S:  g(x)_{\mu\nu} \rightarrow \mathbb{G}(x) \in \mathcal{F}^{n,0}\, . 
\end{equation}
Let $M$ be parallelizable and  choose the set $\tilde{S}=\{V_1, \dots V_n\} \subset\Gamma(EM)$ to be a set of $n$ tetrad fields, then we get that 
\begin{equation}
    \mathbb{G}(x)=\mathbb{1} \qquad \text{ for all } x\in \mathcal{M} \,.
\end{equation}
Then the measure associated with our effective physical model $\Phi(\mathcal{M},g)$ via the local correlation map generated by the (quantum) reference frame $\tilde{S}$ is the Dirac measure $\rho_S^\Phi = \operatorname{Vol}(\mathcal{M}) \delta(\mathbb{1})$.\\
As a result we see that when choosing the reference frame/system to have the same dimension as the fiber space then in the worst case we loose all the information about our physical model, except for the total measure. Essentially the correlation geometry of the entire manifold is collapsed onto a single point/operator. 
\end{example}

\begin{example}\label{ex:two}
    The following example is more closely aligned with the coordinate fields in Kabel et al.\cite{kabel2025quantum}. We take the physical model $\Phi(M,g,\mu_g)$ again to be a compact $n$-dim Riemannian manifold with $\mu_g=\sqrt{|g(x)|}d^n x$. Now instead of vector fields we consider sections of the complex line bundle (the case for real scalar fields is essentially the same). I.e. $EM=\C M $ and $S=\{\psi_1,\dots, \psi_n\}\subset C^\infty(M,\C)$ being $n$ complex-valued scalar fields. 
For the scalar product we choose the Sobolev norm $H^k$ which for $H^1$ is equivalent to
\begin{equation}
    \int_M \left(\overline{\psi_i(x)} \psi_j(x) + g\left(\nabla \overline{\psi_i(x)}, \nabla \psi_j(x)\right)\right) \sqrt{|g(x)|}d^n x
\end{equation}
The bilinear form we choose in this case to be 
\begin{equation} \label{eq:bilintwo}
    \tilde{b}_x(\psi_i,\psi_j)= g\left(\nabla \overline{\psi_i(x)}, \nabla \psi_j(x)\right) \,.
\end{equation}
Then the local correlation map 

\begin{equation}
  (\psi_i,\widetilde{\mathbb{G}}(x)\psi_j)_\H := g_x\left(\nabla \overline{\psi_i(x)}, \nabla \psi_j(x)\right) \qquad \forall \psi_i,\psi_j \in S
\end{equation}
identifies the following form at $x$ with a self-adjoint rank $n$ operator on the Hilbert space $\H=\overline{S}^{H^k}$

\begin{equation}
 F_S:   g_x\left(\nabla \overline{\cdot}, \nabla \cdot \right) \rightarrow \widetilde{\mathbb{G}}(x) \in \mathcal{F}^{n,0}\, . 
\end{equation}
If we restrict to the case of real scalar functions and choose the set $\tilde{S}=\{\psi_1,\dots,\psi_n\}$ accordingly to be a set of $n$ scalar fields then the fact that $\mathcal{M}$ is compact, guarantees, that there exists at least two points $x,y\in \mathcal{M}$ such that $\widetilde{\mathbb{G}}(x) \notin \mathcal{F}^{n,0}_{\text{reg}}$. Because every real scalar function has a minimum and a maximum on $\mathcal{M}$ and at these points the local correlation operator is degenerate. Hence we see also in this case that choosing the dimension of the reference frame/system to small, leads us to loose essential information about our physical model $\Phi(M,g,\mu_g)$.

This example is a good point to remark on a subtly of the construction: 
For the bilinear form \ref{eq:bilintwo} to be well defined, we require that the fields $\psi_i$ are in $C^1(M)$. For a finite dimensional subspace $S$, that is fine, however if $S$ is infinite dimensional some care is required. For $\overline{S}^{H^k}\subset C^1(M)$  the Sobolev embedding theorem requires $k>1+\frac{\dim (M)}{2}$. Alternatively we can work with $H^1$ and introduce a regularization instead by defining the bilinear form as
\begin{equation} \label{eq:bilintwo}
    \tilde{b}^\epsilon_x(\psi_i,\psi_j)= \frac{1}{\operatorname{vol}(B_\epsilon(x))}\int_{B_\epsilon(x)}g\left(\nabla \overline{\psi_i(x)}, \nabla \psi_j(x)\right)\sqrt{|g(x)|}d^n x \,
\end{equation}
where $B_\epsilon(x):=\{y\in M|d(x,y)<\epsilon\}$ and $d(x,y)$ is the Riemannian distance function.
    
\end{example}

In summary, to represent the effective physical model $\Phi$ in terms of its correlation geometry, one needs to make three choices
\begin{enumerate}
    \item Choose a suitable set of sections $S$ over the manifold of the physical model $\Phi(M)$. This is the reference frame/system. 
    \item Define a suitable scalar product on $S$ which fixes in a sense a global frame (or in fact a sense of parallel transport as the orthogonal basis of $\H$ provides a way to identify vectors across different fiber spaces).
    \item Choose a suitable hermitian form $b_x$ to encode local information. The signature of the hermitian form determines the space $\F^{p,q}$ into which we map. Depending on the choice of the scalar product and the desired form of $b_x$, we also need to choose a suitable regularization for the map to be well-defined. 
\end{enumerate} 
The correlation geometry $(\H, \F^{p,q}, \rho_s^\Phi)$ then gives us a description of our effective physical model in terms of the (quantum) reference frame $S$. To be precise, the correlation geometry only retains information about the correlations of the different subsystems of $S$. Hence $(\H_S, \F^{p,q}, \rho_s^\Phi)$ is a purely relational description of the effective physical model $\Phi$ in terms of the reference system.  

The key step now is to consider the correlation geometry $(\H_S, \F^{p,q}, \rho_s^\Phi)$ to be the fundamental physical model giving rise to the effective physical model $\Phi(M,g, \mu)$ as an (approximate) effective description. For an effective physical model to be a viable description of a particular correlation geometry, it must admit a suitable choice of $S$, $\H$, and $b_x$ such that the local correlation map with respect to this particular choice maps to $(\H_S, \F^{p,q}, \rho_s^\Phi)$.  In this sense, as already mentioned, all the descriptors of the effective physical model $\Phi(\mathcal{M}, g, \dots, \mu)$ should be thought of as labels to identify certain special correlation geometries. A priori for a generic measure $\rho$ there is no reason why one should expect a compact description of the correlation geometry in terms of a manifold and a hand full of fields to exist. Similarly, in statistical mechanics, a general statistical distribution does not admit a description in terms of temperature and pressure. 

Continuing the analogy with thermodynamics, the fact that the correlation geometry depends on the reference system $S$ suggests that it should be appended to the descriptors of the effective physical model $\Phi(\mathcal{M}, g, \dots, \mu,S)$ in the same way that the number of particles $N$ is added to the descriptors of a thermodynamic system when understood as arising from the ensemble of statistical mechanics. In the context of CFS in Minkowski space, this suggests that the Dirac sea is not just an auxiliary structure that can be used to build the local correlation map, but it should be treated as part of the effective physical model to the same degree as is done with the number of particles $N$ in thermodynamics. As we argue in Section \ref{sec:unitary} there exists a well-defined manner to compare different effective physical models when the reference system is taken into account, avoiding the obstacles posed by gauge transformations. Now, while we stay here with the language of ``reference frame/system'' for $S$ much as in thermodynamics where the $N$ particles are the ``real'' fundamental degrees of freedom, the system $S$ should be treated as the fundamental degrees of freedom and the correlation geometry, in particular the measure, describes the dynamics of a many particle quantum system in the same way the density function on phase space describes the dynamics of a many particle classical system. This point of view was already developed to some degree in \cite{fischer2025causal}. In contrast to the statistical description of classical thermodynamics, the correlation geometry, however, does not require a background phase space to operate in, opening the door to a unified description of our physical world.

With this analogy at hand, it is easy to see why a ``superposition'' of effective physical models as argued for in \cite{kabel2025quantum}, or of spacetimes more general as argued for in, e.g., \cite{belenchia2018quantum,zych2019bell,christodoulou2019possibility,anastopoulos2020quantum,giacomini2022quantum,foo2023quantum}, is not necessarily a sensible thing to do in our framework, beyond the linearized/weak-field regime. For example, it is not clear to us to what degree one can make sense of a ``superposition'' of temperatures in thermodynamics. Of course, one can add the statistical distributions corresponding to different temperatures. However, this new system would exhibit dynamics that would be distinctly different from two isolated thermodynamic systems at different temperatures. The newly established ``mixed'' system would evolve towards a new equilibrium distribution at a different temperature. The naive superposition would therefore correspond to a statistical distribution that is out of equilibrium. These are of course interesting systems, but in contrast to quantum systems in superposition they give rise to qualitatively new dynamics. In Section \ref{sec:unitary} we will argue that the situation for correlations geometries is indeed quite similar and hence a ``superposition'' of spacetimes is, in general, not a well-defined operation\footnote{In \cite{fischer2025causal} we argued, that the total \ac{cfs} spacetime can be considered as a superposition of sub-system spacetimes. To avoid confusion it is important to note here, that these sub-system spacetimes refer to the spacetime notion at the fundamental level. It is by no means certain (and in general not true) that these sub-system spacetimes admit a description in terms of a classical spacetime as an effective physical model.}.

\subsection{Encoding Matter Fields}
In the above Examples \ref{ex:one} and \ref{ex:two} we showed that the local correlation map encodes a local bilinear form $b_x$ as a self adjoint operator on a globally defined Hilbert space $\H$. The procedure in section \ref{sec:encoding} allows us to encode (aspects of) the structure of the manifold $\mathcal{M}$ of our effective physical model $\Phi$ and its associated volume form $\mu$. In the examples, the elements in $\F^{p,q}$ encode aspects of the metric.

In Section \ref{sec:encoding} we highlighted the crucial role that the choice of the reference system $S$ plays in the construction of the local correlation map and hence in the corresponding correlation geometry. In this section, we will explain how this ``freedom'' of choice can be used to encode all physical fields $A_i$ used as descriptors of an effective physical model $\Phi(\mathcal{M}, g, \mu, A_i, \dots)$.

To encode physical fields of the effective physical model $\Phi(\mathcal{M}, g, \mu, A_i, \dots)$ in the correlation geometry, we have to make the choice of the reference system $S\subset\Gamma(EM)$ dependent on these fields. The way to do this is to consider an operator
\begin{equation}
    \mathcal{D}[g, A_i, \dots] : \qquad\Gamma(EM) \rightarrow\Gamma(EM)\, .
\end{equation}
We then consider $S[g, A, \dots]\subset \ker(\mathcal{D}[g, A, \dots])\subset \Gamma(EM) $ and the Hilbert space again the closure of $S$ with respect to a suitable scalar product.
\begin{example}
    In the context of \ac{cfs} the fiber bundle of choice is  the spin bundle over the manifold $EM=SM$ and the operator $ \mathcal{D}[g, A, \dots]$ is the Dirac operator. 
    \begin{enumerate}
        \item For an effective physical model $\Phi(\mathcal{M}, g, \mu_g)$ that describes a vacuum spacetime, the equation of choice is the massive Dirac equation
        \begin{equation}\label{eq:dirac}
           ( \slashed{D} +m)\psi(x)=0
        \end{equation}
        where $ \slashed{D}$ is the Dirac operator on $SM$. To model Minkowski vacuum, e.g., one chooses $S[\mathcal{M},g]$ to be the negative energy solutions of \eqref{eq:dirac}, see, e.g., \cite{cfs,current} for details.
        \item For an effective physical model $\Phi(\mathcal{M}, g, \mu_g, A)$ describing a spacetime with an electromagnetic field $F=dA$ the equation of choice is the minimally coupled massive Dirac equation
        \begin{equation}\label{eq:dirac}
           ( \slashed{D} -iq\slashed{A} +m)\psi(x)=0
        \end{equation}
        where $ \slashed{D}$ is again the Dirac operator on $SM$ and $A$ is the vector potential of the Maxwell field. 
    \end{enumerate}
\end{example}
As a consequence of the dependence of the subspace $S[g, A, \dots]$ on the physical fields, the local correlation map 
\begin{align}
 \hphantom{xxxxx}   F[g, A, \dots]:&&  \mathcal{M} &&\longrightarrow &&&\F^{p,q} \hphantom{xxxxx}  \\
    && x &&  && & \mathbb{F}[g, A, \dots](x)\hphantom{xxxxx} 
\end{align}
now also depends on these fields and as a consequence also on the push forward measure $\rho_S^\phi[g, A_\mu, \dots]$ representing the correlation geometry of this effective physical model on the fundamental level.

\subsubsection{Comparing Physical Models - Equivalence, Symmetries and Gauge Invariance} \label{sec:unitary}
Having introduced effective physical models $\Phi$ and their descriptors in terms of physical fields $\Phi[\mathcal{M}, g, A, \dots]$ and how they are mapped to an associated correlation geometry $(\H_S, \F^{p,q}, \rho_S^\phi[g, A, \dots])$, we are now able to compare different effective physical models $\Phi_i[\mathcal{M}_i,g_i, A_i, \dots, S]$. We start with the following definition.
\begin{definition}[Unitary Equivalence of Fundemantal Physical Models]\label{def:equivalence}
    Let $(\H_1, \F^{p_1,q_1}, \rho_{S, 1}^\phi[g_1, A_1, \dots])$ and $(\H_2, \F^{p_2,q_2}, \rho_{S, 2}^\phi[g_2, A_2, \dots])$ be two a priori distinct fundamental physical models with $\dim(\H_1)\geq\dim(\H_2)$. We say they are unitarily equivalent, if upon isometric embedding 
    \begin{equation}
        \iota(\H_2)\hookrightarrow H_1 
    \end{equation}
    with the embedding induced by a trivial extension on $\operatorname{im}(\iota)^\perp$
    \begin{equation}
        \iota^*(\F^{p_2,q_2}(\H_2))\hookrightarrow \F^{p_2,q_2}(\H_1)
    \end{equation}
    there exists a unitary transformation $U$ on $\H_1$ such that
    \begin{equation}
        \rho_{S, 1}^\phi[g, A, \dots](UxU^{-1}) = \rho_{S, 2}^\phi[g, A, \dots](\iota^{-1}(x))\,. 
    \end{equation}
\end{definition}
\begin{remark}
 In general, it only makes sense to compare fundamental physical models for which the local correlation operators $x\in\F^{p_i,q_i}$ have the same number of positive and negative eigenvalues $p_i=p_j$ and $q_i=q_j$. Now, a priori, two fundamental physical models can still be unitarily equivalent even if  $p_i\neq p_j$ or $q_i\neq q_j$ or both. However, in this case either for one measure or for the other or for both, we have to have that $\operatorname{supp}(\rho_{S, i}^\phi[g_i, A_i, \dots])\cap \F^{p_i,q_i}_{\text{reg}}=\emptyset$. 
\end{remark}

\begin{remark}
    The unitary transformation in Definition \ref{def:equivalence} accounts for the fact, that the embedding $ \iota(\H_2)\hookrightarrow H_1 $ is only fixed up to unitary transformations. Whether or not two fundamental physical models are equivalent should not dependent on this choice of embedding. 
\end{remark}
We can now define what we mean for two physical models to be the same. 

\begin{definition}[Equivalence of Effective Physical Models]\label{def:sameness}
Assume that the effective physical models $ \Phi_1[\mathcal{M}_1,g_1, A_1, \dots,S_1]$ and $ \Phi_2[\mathcal{M}_2,g_2, A_2, \dots,S_2]$ correspond to the fundamental physical models
     $(\H_1, \F^{p_1,q_1}, \rho_{S_1}^\phi[g, A, \dots])$ and $(\H_2, \F^{p_2,q_2}, \rho_{S_2}^\phi[g, A, \dots])$ 
      We say that the two effective physical models are indistinguishable 
      \begin{equation}
          \Phi_1[\mathcal{M}_1,g_1, A_1, \dots,S_1]\sim \Phi_2[\mathcal{M}_2,g_2, A_2, \dots,S_2] \, ,
      \end{equation}if their fundamental physical models are unitarily equivalent.
\end{definition}
We will now show that this notion of equivalence of effective physical models is flexible enough to encode both classical gauge transformations and diffeomorphisms. It might be worth pointing out at this point that the representation of an effective physical model $\Phi$ in terms of its fundamental representative $(\mathcal{H}_S, \F^{p,q},\rho_{S}^\phi)$ is trivially coordinate invariant. We will now start by considering the gauge freedom of electro magnetism as the most straightforward example to illustrate the principle of unitary equivalence. 
\begin{example}\label{ex:three}
    Let $\Phi_1(\mathcal{M},g,A, \mu_g)$ and $\Phi_2(\mathcal{M},g,A,\chi,\mu_g)$ be two effective physical models with $\Phi_2(\mathcal{M},g,A,\chi,\mu_g):=\Phi_1(\mathcal{M},g,A+d\chi, \mu_g)$. 
Let the reference system $S_1:=\{\psi_1,\dots,\psi_m\}$ be a subset of $\ker\left( \slashed{D} -iq\slashed{A} +m\right)$, i.e. solutions to the Dirac equation with minimal coupling to the electromagnetic potential, and let the local hermitian form be $C^\infty(\mathcal{M},\mathbb{C})$-linear /anti-linear, i.e. for all $f,g\in C^\infty(\mathcal{M},\mathbb{C})$ we have 
\begin{equation}
    b_x(f\psi_i,g \psi_j)= \overline{f}g  b_x(\psi_i,\psi_j)\, \footnote{For example $b_x(\psi_i, \psi_j)=\overline{\psi_i(x)}\psi_j(x)$ would satisfy this condition.} .
\end{equation}
Let the second reference system  be $S_2:=\{e^{-iq\chi}\psi_1,\dots,e^{-iq\chi}\psi_m\}= e^{-iq\chi} S_1$ which is now a subset of $\ker\left( \slashed{D} -iq(\slashed{A}+\slashed{d}\chi) +m\right)$ then we have that
\begin{equation}
    b_x(e^{-iq\chi}\psi_i, e^{-iq\chi}\psi_j)=b_x(\psi_i, \psi_j)\; \; \forall i,j\in [1,m];\;\; \forall  x\in \mathcal{M} \, .
\end{equation}
As a result of the identification $S_2=e^{-iq\chi} S_1$ we have that 
\begin{equation}
    \rho_{S_1}^{\phi_1}[g, A,\mu_g](\mathbb{F}(x)) =\rho_{S_2}^{\phi_2}[g, A, \chi,\mu_g](\mathbb{F}(x))\qquad \forall x\in M
\end{equation}
and according to our Definition \ref{def:sameness} the two effective physical models are equivalent, $\Phi_1(\mathcal{M},g,A, \mu_g,S_1)\sim \Phi_2(\mathcal{M},g,A,\chi,\mu_g,S_2)$, in agreement with conventional wisdom.
\end{example}
Being able to account for gauge transformations of vector fields is one thing, but that is not all. More importantly, as the following example shows, the setup can also be applied to diffeomorphisms. 
\begin{example}\label{ex:four}
    Let there be two, apriori distinct, effective physical models $\Phi_1(\mathcal{M},g_{\mathcal{M}},A_{\mathcal{M},i},\dots, \mu_{g_\mathcal{M}})$ and $\Phi_2(\mathcal{M}, h,\mathcal{N},g_{\mathcal{N}},A_{\mathcal{N},i},\dots,\mu_{g_\mathcal{N}})$ with the desciptor $h: \mathcal{N}\rightarrow \mathcal{M} $ being a diffeomorphism. Furthermore assume that 
    \begin{align}
        \Phi_2&(\mathcal{M}, h,\mathcal{N},g_{\mathcal{N}},A_{\mathcal{N},i},\dots,\mu_{g_\mathcal{N}}):=\\
        &\Phi_1(\mathcal{M}= h(\mathcal{N}),g_{\mathcal{M}}=h^*(g_{\mathcal{N}}),A_{\mathcal{M},i}=h^*A_{\mathcal{N},i},\dots, \mu_{g_\mathcal{M}}= \mu_{g_\mathcal{N}}\circ h ) \, .
    \end{align}
    Here $h^*$ denotes the suitable pushforward map for the various descriptors of the effective physical model $\Phi_1$.
    Now let $S_1\subset E\mathcal{M}$ be the reference frame for the effective physical model $\Phi_1$ and let $S_2\subset \tilde{E} \mathcal{N}$ be a reference frame for the effective physical model $\Phi_2$. Assume that the reference frame of the effective physical model $\Phi_1$ is the suitable pushforward of the reference system of the effective physical model $\Phi_2$, $S_1=h^*S_2$\footnote{Here we have to assume that the fiber spaces $E_x 	\cong \tilde{E}_p $ are isomorphic.}. Furthermore assume that the local hermitian forms $b_x^\mathcal{M}=h^*b^\mathcal{N}_{h(x)}$  and the scalar products $(\cdot|\cdot)_\mathcal{M} = h^*\left(  (\cdot|\cdot)_{\mathcal{N}} \right) $ are related accordingly. As a result of the identification $S_1=h^*S_2$ we then  have that
    \begin{equation}
    \rho_{S_1}^{\phi_1}[g_{\mathcal{M}},A_{\mathcal{M},i},\dots,\mu_{g_\mathcal{M}}](\mathbb{F}_\mathcal{M}(h(x))) =\rho_{S_2}^{\phi_2}[g_{\mathcal{N}},A_{\mathcal{N},i},\dots,\mu_{g_\mathcal{N}}](\mathbb{F}_\mathcal{N}(x))
\end{equation}
    and hence, according to Definition \ref{def:sameness}, the two effective physical models $\Phi_1(\mathcal{M},g_{\mathcal{M}},A_{\mathcal{M},i},\dots, \mu_{g_\mathcal{M}},S_1)\sim \Phi_2(\mathcal{M}, h,\mathcal{N},g_{\mathcal{N}},A_{\mathcal{N},i},\dots,\mu_{g_\mathcal{N}},S_2)$\footnote{Given the fact that one needs to assume that the local hermitian forms $b_x=h^*\tilde{b}_{h(x)}$  and the scalar products $(\cdot|\cdot)_\H = h^*\left(  (\cdot|\cdot)_{\tilde{\H}} \right) $ are related in a suitable manner, one could argue that they too should appended to the descriptors of the effective physical modeland thus the equivalence would read: $\Phi_1(\mathcal{M},g_{\mathcal{M}},A_{\mathcal{M},i},\dots, \mu_{g_\mathcal{M}},S_1,(\cdot|\cdot)_\mathcal{M}, b_x^\mathcal{M})\sim \Phi_2(\mathcal{M}, h,\mathcal{N},g_{\mathcal{N}},A_{\mathcal{N},i},\dots,\mu_{g_\mathcal{N}},S_2,(\cdot|\cdot)_\mathcal{N}, b_x^\mathcal{N})$} are indeed equivalent in agreement with conventional wisdom that effective physical models that are related by a diffeomorphism are equivalent with regards to their physical content. 
    
    Note however, that in the context of correlation geometry not only the models themselves have to be related by a diffeomorphism, but also the reference frame and the structures required to encode their relational information, i.e. the scalar products and the local hermitian forms.
\end{example}

These examples motivate the following definition. 
\begin{definition}[Gauge Transformations]
    We call any mathematical object in the effective physical model $\Phi_2$ that is redundant, in the sense that there exists an equivalent effective physical model $\Phi_1\sim \Phi_2$ without these mathematical objects, a gauge transformation. 
\end{definition}

\begin{remark}
    The scalar field $\chi$ in Example \ref{ex:three} and the pair $\{h^{-1},\mathcal{N}\}$ in Example \ref{ex:four} are gauge transformations in this sense. In Example \ref{ex:four} one could equally well add the objects $\{h^{-1},\mathcal{N}\}$ to the descriptors of the effective physical model $\Phi_1$ and remove $\{h,\mathcal{M}\}$ from the descriptors of the effective physical model $\Phi_2$. In this case, we would get the same result, except that now $\{h^{-1},\mathcal{M}\}$ in $\Phi_1$ are redundant instead of $\{h,\mathcal{N}\}$ in $\Phi_2$. Neither gauge is preferable in an absolute sense, as it should be. 
\end{remark}

While the gauge transformations are redundancies in the descriptors of the physical models $\Phi$ the notion of a unitary equivalence motivates the following definition of a symmetry for a correlation geometry. For \ac{cfs} this was introduced and studied in \cite{osymm,sigsymm,lagrange,pmt,Finster2024}.
\begin{definition}[Symmetry for Correlation Geometry]\label{def:symmetry}
    Let $(\mathcal{H}, \mathcal{F}^{p,q}, \rho)$ be a specific fundamental physical model. We say that a unitary transformation $U$ is a symmetry of this model if the measure is invariant under the action of $U$
    \begin{equation}
        \rho(\Omega)= \rho(U\Omega U^{-1})\qquad \text{ for all measurable sets } \qquad \Omega \subset \mathcal{F}^{p,q} \, .
    \end{equation}
\end{definition}
A conjecture of current interest related  to this definition is as follows.
\begin{conjecture}
    Every effective physical model that features an isometry that leaves all fields invariant, and is mapped by a reference system $S$ satisfying an equation invariant under the isometry, gives rise to a correlation geometry with a related symmetry according to Definition \ref{def:symmetry}.
\end{conjecture}

\section{Conclusion}
A key insight in this paper is that correlation geometry comes with gauge invariance built in. Crucially, this notion of gauge transformations also covers diffeomorphism, which in the past has proven a rather difficult obstacle for many approaches to unification/quantum gravity. 

When it comes to the ideas regarding spacetime superpositions as discussed in \cite{belenchia2018quantum,zych2019bell,christodoulou2019possibility,anastopoulos2020quantum,giacomini2022quantum,foo2023quantum,kabel2025quantum} we here get the following picture which we will argue is much more in line with thermodynamics than quantum theory:\\
Suppose that we have two effective physical models $\Phi_1$ and $\Phi_2$ with their corresponding fundamental physical models $(\H_1, \F^{p,q}, \rho_S^{\phi_1})$ and $(\H_2, \F^{p,q}, \rho_S^{\phi_2})$ formulated in terms of correlation geometry. Upon identification of the Hilbert spaces $\H_1\hookrightarrow\H_2$ one is of course free to consider any convex combinations of the two measures as a new fundamental physical model $(\H_2, \F^{p,q}, \tau\rho_S^{\phi_2}+(1-\tau)U\rho_S^{\phi_1}U^{-1} )$. Here again we have the freedom to pick an arbitrary unitary transformation $U$ on $\H_2$ to account for the fact that the embedding  $\H_1\hookrightarrow\H_2$ is only determined up to unitary transformation.

Now, while $(\H_2, \F^{p,q}, \tau\rho_S^{\phi_2}+(1-\tau)U\rho_S^{\phi_1}U^{-1} )$ is a perfectly valid correlation geometry in general, one would not expect that it can readily be mapped via a local correlation map to a simple effective physical model $\Phi_3$ with comparably few descriptors as the models $\Phi_1$ and $\Phi_2$. 

The situation is  similar to the situation in thermodynamics, where for a gas in equilibrium the underlying statistical distribution is given in terms of a handful of macroscopic parameters. Now, if we have two gasses at equilibrium at different temperatures with probability distributions $P_1(T_1,p_1,\dots)$ and $P_2(T_2,p_2,\dots)$, combining them to one system by writing a ``superposition'' of their temperatures is not a well-defined operation. However, what is well defined is the operation of taking a linear, or convex, combination of the two probability distributions
\begin{equation}
    P= \tau P_1(T_1,p_1,\dots) + (1-\tau)P_2(T_2,p_2,\dots) \, .
\end{equation}

Now, again, $P$ is a perfectly valid probability distribution for statistical mechanics. However, it no longer admits a simple description in terms of a handful of equilibrium parameters. In fact, what we have created here is a state $P$ that is \emph{out of equilibrium}. In statistical mechanics, of course, we know what happens next when we create a system with a probability distribution $P$: It evolves towards a new equilibrium state $\tilde{P}$ which can then again be described in terms of a handful of macroscopic parameters $\tilde{P}(T_3,p_3,\dots)$. However, this is in \emph{stark} contrast to quantum mechanical superposition, where the new state evolves happily ever after according to the same dynamical laws as the two states independently. 

Whether in correlation geometry we can make sense of $(\H_2, \F^{p,q}, \tau\rho_S^{\phi_2}+(1-\tau)U\rho_S^{\phi_1}U^{-1} )$ as an out of equilibrium system as well is unclear at the moment.  The nature of the construction in any case suggests rather an alignment with the thermodynamic picture rather than the quantum mechanical one, and therefore one should expect a qualitatively different dynamics for such systems. 

\thanks{{{\em{Acknowledgments:}} 
We thank \v{C}aslav Brukner, Flaminia Giacomini, Leon Loveridge, Felix Finster, and Antonino Marciano for helpful comments on the draft. }

    \printbibliography
\end{document}